%% file: ms.tex
\documentclass{emulateapj}

\def\chandra{{\it Chandra}}
\def\xsix{X1624$-$490}
\def\sco{Sco~X$-$1}
\def\x1702{X1702$-$429}

\begin{document}
\title{Chandra HRC Localization of the Low Mass X-ray Binaries X1624$-$490 and 
X1702$-$429: The Infrared Counterparts} 
\author{Stefanie Wachter\altaffilmark{1}}  
\affil{Spitzer Science Center, California Institute of Technology, 
MS 220-6, Pasadena, CA 91125}
\email{wachter@ipac.caltech.edu}

\author{Joseph W. Wellhouse}
\affil{Harvey Mudd College, 340 East Foothill Blvd., Claremont, CA 91711}

\author{Sandeep K. Patel}
\affil{Universities Space Research Association, NSSTC, SD-50, 320 Sparkman Drive, 
Huntsville, AL 35805}
\email{Sandeep.Patel@msfc.nasa.gov}

\author{Alan P. Smale\altaffilmark{2}}
\affil{Laboratory for High Energy Astrophysics/USRA, Code 662, NASA/GSFC,
Greenbelt, MD 20771}
\email{alan@osiris.gsfc.nasa.gov}

\author{Joao F. Alves}
\affil{European Southern Observatory, Karl-Schwarzschild-Strasse 2, D-85748 Garching bei M\H{u}nchen, Germany }
\email{jalves@eso.org}

\author{Patrice Bouchet}
\affil{Cerro Tololo Inter-American Observatory, National Optical Astronomy 
Observatory\altaffilmark{3}, Casilla 603, La Serena, Chile}
\email{pbouchet@noao.edu}

\altaffiltext{1}{also with Eureka Scientific Inc., 2452 Delmer Street, Suite 100, Oakland, CA 94602}
\altaffiltext{2}{Now at Office of Space Science, NASA Headquarters, Washington DC 20546} 
\altaffiltext{3}{Operated by the Association of Universities for Research in
Astronomy, Inc., under cooperative agreement with the National Science
Foundation.}

\begin{abstract}

We report on the precise localization of the low mass X-ray binaries 
\xsix\ and \x1702\ with the \chandra\ HRC-I. We determine the best positions to be 
16$^h$28$^m$02\fs825 $-49^{\circ}$11\arcmin 54\farcs61 (J2000) and
17$^h$06$^m$15\fs314 $-43^{\circ}$02\arcmin 08\farcs69 (J2000) 
for \xsix\ and \x1702, respectively, 
with the nominal \chandra\ positional uncertainty of 0.6\arcsec.
We also obtained deep IR observations
of the fields of these sources in an effort to identify the IR counterparts. 
A single, faint ($Ks=18.3 \pm 0.1$) source is visible inside the \chandra\ error circle of 
\xsix\, and we
propose this source as its IR counterpart. For \x1702, a $Ks= 16.5 \pm 0.07$ source is visible at the edge 
of the \chandra\ error circle. The brightness of both counterpart candidates is comparable to 
that of other low mass X-ray binary IR counterparts when corrected for extinction and distance. 

\end{abstract}

\keywords{stars:neutron --- X-rays:binaries --- X-rays:individual(4U~1624$-$490, 4U~1702$-$429)}

\section{Introduction}

\xsix\ belongs to the group of low mass X-ray binaries (LMXBs) 
that exhibit periodic intensity dips
in their X-ray light curves.
It is generally accepted that these so-called ``dippers'' are high inclination
systems and that the dips are due to occultations of the
central X-ray source by a thickened region of the accretion disk rim where the
gas stream from the companion impacts upon the outer disk. 
An orbital period of $\sim21$~h for \xsix\ was first reported by 
\citet{wat85} and recently refined to 20.89~h by \citet{smale01}. 
\xsix\ is one of the most extreme systems among the dippers; its 
persistent X-ray emission is the brightest ($6\times 10^{37}$erg s$^{-1}$), 
its dip profiles seemingly the most erratic, and its 20.89-hour orbital period is the 
longest. The orbital period is $\sim$ 5--25 times longer than those of 
the other dipping sources, 
corresponding to a much greater stellar separation and a larger accretion disk radius. 
Dipping is deep, $\sim$75\% in the 1--10 keV band, and can at 
times reach a stable lower level \citep{cbc95} 
suggesting that one spectral component is entirely removed. The source also 
exhibits strong flaring in which the X-ray flux can increase by $\sim$30\% 
over timescales of a few thousand seconds. Most recently, \citet{parmar02} 
discovered narrow absorption features in the \chandra\ X-ray spectrum of \xsix.
Despite the wealth of X-ray observations, the optical/IR counterpart of
\xsix\ has not yet been identified. In fact, it is the only
one of the known dipping sources without an identified counterpart.

\x1702 was first discovered as a burst source with OSO 8 \citep{swank76}. 
A persistent X-ray counterpart for this burster was subsequently suggested by 
\citet{lewin79}. This identification was supported by observations
from \citet{maki82} who also found variability in the burst activity of the 
source. \citet{ooster91} classified \x1702\ as an Atoll source based on EXOSAT observations. 
\x1702\ is remarkable because it is one of the few X-ray bursters
that shows kHz quasi-periodic oscillations (QPOs) 
in both its persistent X-ray emission and during type I X-ray bursts, 
the latter of which might directly indicate the spin period of the 
neutron star (\citealt{stroh99}; \citealt{mark99}).  

We report here on the precise localization of \xsix\ and \x1702\ with
\chandra\ and the subsequent identification of infrared counterpart candidates for both sources.

\section{Observations}

\subsection{X-ray}
We observed \xsix\ (Obs ID 2693) on 2002 May 30 and \x1702\ (Obs ID 3791) on 2003 June 19 with 
the \chandra\ HRC-I for 1 ksec each. 
CIAO v.3.0 and CALDB v2.26 were used in all
data analysis and processing tasks. There were no known 
aspect corrections for these data sets. 
We consequently assume an error radius of 0.6\arcsec\ (90\% confidence) for the derived positions of the
X-ray sources from these data according to the information provided at 
http://cxc.harvard.edu/cal/ASPECT/celmon/.

We searched for X-ray sources using both {\it wavdetect} and our own source finding algorithm. 
In the \xsix\ data set, a single bright source is detected at the center of the 
30\arcmin $\times$30\arcmin\ field. The best position is 
16$^h$28$^m$02\fs825 $-49^{\circ}$11\arcmin 54\farcs61 (J2000) with the
0.6\arcsec\ uncertainty mentioned above.
The ROSAT All Sky Survey (RASS) sources (1RXS J162622.5-491419 and 1RXS J162824.6-492510) 
that we hoped to use 
to improve the global astrometry were not detected. Both are very weak sources in the 
RASS, and the \chandra\ non-detection implies that they are either not real or transient sources.
However, we do find a weak (0.009 counts sec$^{-1}$, S/N$\sim 3$) source in our \xsix\ \chandra\ data at  
16$^h$27$^m$36\fs262 $-49^{\circ}$13\arcmin 03\farcs31 (J2000). Its position matches that of the 
bright 2MASS source 16273627-4913042 at 
16$^h$27$^m$36\fs271 $-49^{\circ}$13\arcmin 04\farcs26 (J2000). 
The colors and magnitudes of the 2MASS source ($J=8.936\pm 0.023$, $H=8.245\pm 0.044$, 
$K=8.057\pm 0.024$) are 
consistent with those of a K4-K5 giant at $\sim 2$~kpc based on the data 
by \citet{bb88} transformed
to the 2MASS photometric system with the relations by \citet{carpenter01}. 
If this association is real, the offset between the \chandra\ and 2MASS astrometric frame is 
no larger than 0.9\arcsec\ and predominantly in the DEC coordinate. 

For \xsix\, we determine an HRC-I count rate of 4.25$\pm 0.07$ counts sec$^{-1}$, which falls in the 
low end of the range predicted using the non-dip, non-flare spectral parameters in the literature 
(e.g. \citealt{cbc95}; \citealt{bc2000}; \citealt{smale01}). According to the source ephemeris \citep{smale01}, our 
\chandra\ observation covered phases 0.31--0.34, thus well away from the intervals where dipping activity 
would be expected.

In the \x1702\ data set, the X-ray binary is the only source detected, with a count rate of
10.46$\pm 0.10$ counts sec$^{-1}$. This count rate is consistent with the rate we predicted for the \chandra\ 
HRC-I based on the published flux levels and spectra of this source. Our best localization 
gives 17$^h$06$^m$15\fs314 $-43^{\circ}$02\arcmin 08\farcs69 (J2000). Again, the weak RASS source
1RXS J170721.9-431436 we had hoped to use for refining the match between the X-ray and 
IR astrometric frames is not detected. However, the source would have been at the very edge
of the FOV given the roll angle of our \chandra\ observation, so the non-detection is not conclusive. 

\subsection{Infrared}

Observations of the field of \xsix\ were obtained with the CTIO 1.5m and CIRIM on 1997 
April 26 - 28 in the $Ks$ and $J$ filter. A log of all infrared observations of 
the fields of \xsix\ and \x1702\ is provided in Table~\ref{t-log}. 
The data were dark subtracted and flatfielded 
in the standard manner using IRAF. A normalized sky was created from the median of the object
data, scaled to the median of each image and subtracted. Finally the images were shifted and 
added into a combined final frame excluding bad pixels via a mask. Photometric and astrometric 
calibration of this frame was accomplished through a comparison to 2MASS.  
No IR source is visible in these data at the \chandra\ position of 
\xsix\ down to a magnitude limit of $Ks\approx 18.0$. 

We consequently obtained deeper observations
with the ESO NTT and SOFI on 2003 June 19. SOFI was used in the large field mode which 
provides a 0.29\arcsec\ pixel$^{-1}$ scale. 
The observations were obtained under $\sim 0.8$\arcsec\
seeing conditions. The data were reduced in a similar manner to the 
CTIO 1.5m observations described above, with the addition of the recommended crosstalk 
correction\footnote{see SOFI Manual, 5.1.4 (August 2003 version), and 
http://www.eso.org/$\sim$gfinger/hawaii\_1Kx1K/crosstalk\_rock/crosstalk.html}. 
We derived an astrometric solution for the combined frame through a comparison 
against 82 2MASS stars, using the IRAF task {\it ccmap}. The fit has an accuracy of 
0.05\arcsec\ in each coordinate. Hence, the intrinsic precision of the \chandra\ 
position is still the dominant source of uncertainty in the localization of the 
source. 
We cannot improve the astrometric precision over the nominal 0.6\arcsec\ without 
several other X-ray sources that
can be matched to IR sources in the field. We judge the single, weak, additional source 
detected in our \chandra\ image to be of insufficient quality as a basis for an 
overall shift in the astrometric solution. Instead, we view it as a large scale check 
against a significant systematic offset between the two coordinate systems, as mentioned 
above. 

IR observations of the field of \x1702\ were obtained with the CTIO 1.5m and 4m telescopes. 
Again, the data were reduced in a similar manner to those of \xsix. As is the case for 
\xsix, the CTIO 1.5m data were of insufficient depth to reveal any counterpart candidates
for \x1702. Our most recent CTIO 4m ISPI observations, however, show a counterpart candidate 
at the edge of the \chandra\ X-ray error circle. ISPI has a pixel scale of 
0.3\arcsec\ pixel$^{-1}$
and the seeing conditions during the observations varied between 1.0-1.2\arcsec. 
The astrometric solution for the 4m frame was obtained
from comparison to 65 2MASS stars with residuals of 0.04\arcsec\ in each coordinate.
We derived photometric calibration transformations for both the \xsix\ ESO and \x1702\ 
CTIO observations based on a
comparison to the 2MASS measurements, neglecting any color terms since we only have data
in one filter. The formal rms error in the transformation fit for \xsix\ and \x1702\ was 0.06 and 0.04, 
respectively.

\section{Results}

\begin{figure}[tb]
\plotone{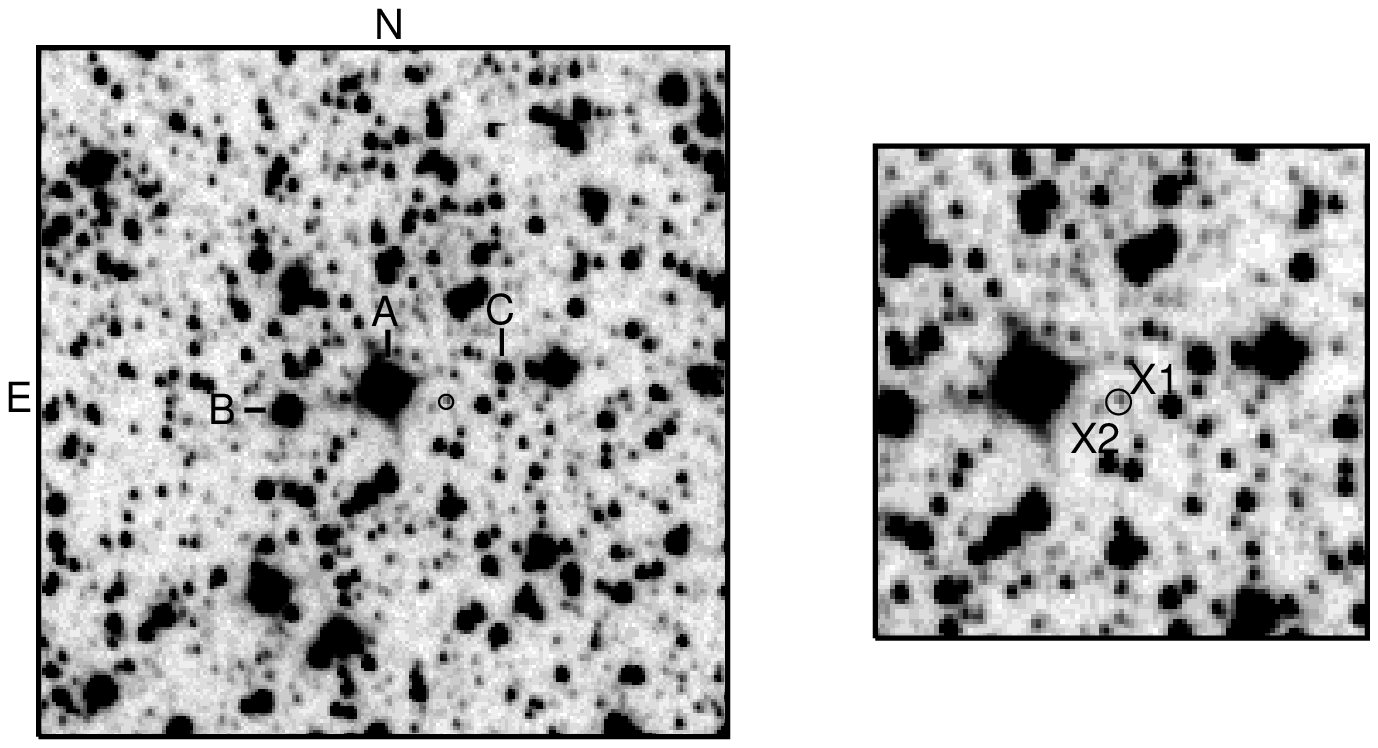}
\figcaption{$Ks$ band image of the field of \xsix, a 1\arcmin $\times$1\arcmin\ field of
view on the left and a 30\arcsec$ \times$30\arcsec\ close-up on the right. 
The best \chandra\ position of 
16$^h$28$^m$02\fs825 $-49^{\circ}$11\arcmin 54\farcs61 (J2000)
with an error
circle of 0.6\arcsec\ radius is shown. A faint source (X1) is visible
inside the error circle.
\label{f-fc1} }
\end{figure}

\begin{figure}[tb]
\plotone{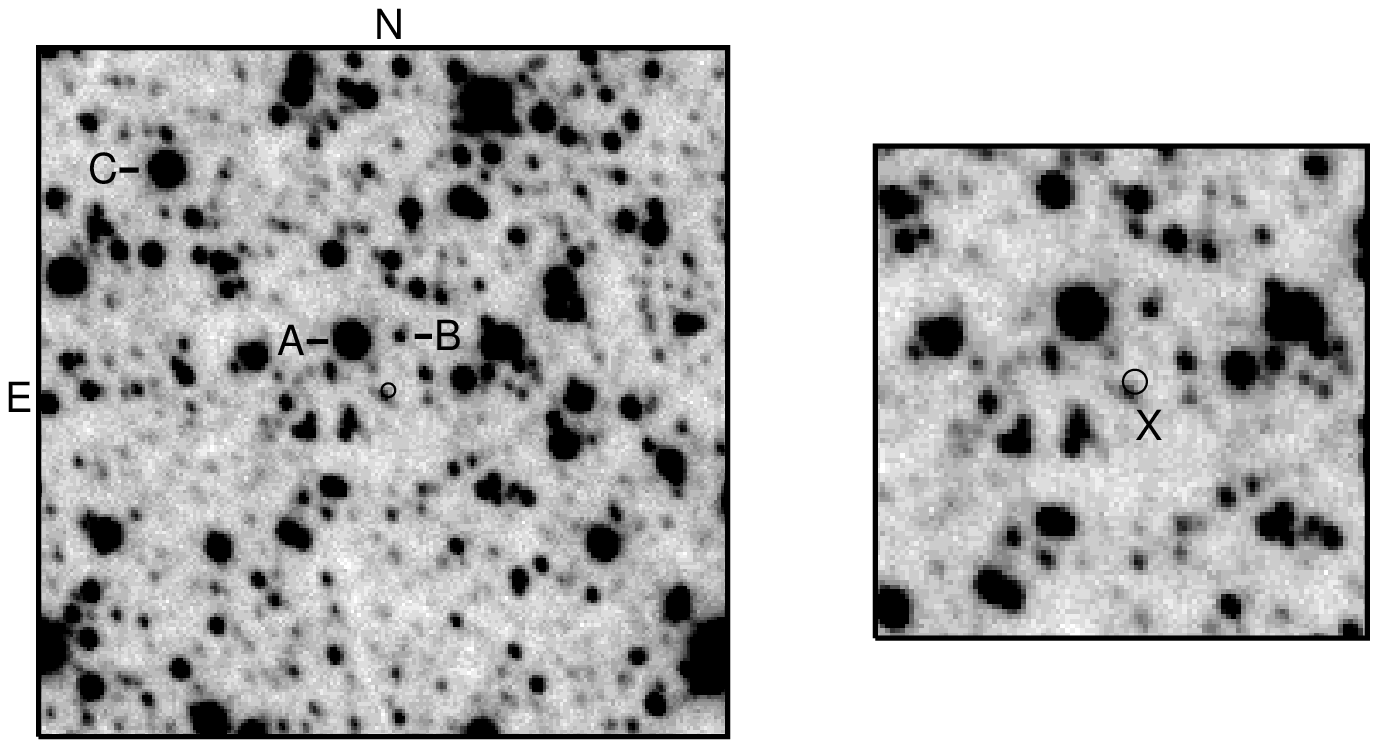}
\figcaption{$Ks$ band image of the field of \x1702, a 1\arcmin $\times$1\arcmin\ field of
view on the left and a 30\arcsec$\times$30\arcsec\ close-up on the right. 
The best \chandra\ position of
17$^h$06$^m$15\fs314 $-43^{\circ}$02\arcmin 08\farcs69 (J2000)
with an error circle of 0.6\arcsec\ radius is shown. A faint source (X) is visible
at the edge of the error circle.
\label{f-fc2} }
\end{figure}

A 1\arcmin $\times$1\arcmin\ portion of our combined $Ks$ frames as well as a 
30\arcsec $\times 30$\arcsec\ close-up around the \chandra\ position of \xsix\ and \x1702\
are shown in Figure~\ref{f-fc1} and Figure~\ref{f-fc2}, respectively. 
The \chandra\ X-ray error circles and several comparison stars are 
marked. Photometry was performed using {\it Daophot} \citep{stet92} and the measurements for all sources 
indicated in the finding charts for both \xsix\ and \x1702\ are summarized in Table~\ref{t-phot}.
We quote the 2MASS magnitudes for those sources detected with the survey and our own 
measurements for the fainter sources only resolved with our dedicated observations. 
Since each potential counterpart is only detected in the combined image of the deepest 
observations, we are unable to place constraints on the variability of these IR counterparts.

\subsection{\xsix}

A single, faint source (X1) is visible inside the \chandra\ error circle, and we
propose this source as the IR counterpart to \xsix.
X1 has a magnitude of $Ks=18.3 \pm 0.1$. X2, the next closest source
to the \chandra\ X-ray error circle, is 1.5\arcsec\ away from X1 with a magnitude of $Ks = 18.6 \pm 0.2$.
In order to evaluate whether this $Ks$ band brightness of X1 is reasonable for a \xsix\ counterpart, we 
compare it to that of \sco. The two systems have similar orbital periods (\sco: 18.9~h; \xsix: 20.9~h) and
X-ray luminosities \citep{christ97} and hence might contain a similarly sized accretion disk and mass donor 
producing the IR emission. One difference between the two systems is the inclination, with \sco\
thought to be close to face-on while \xsix\, as a dipper, is close to edge-on. 
Table~\ref{t-res} lists the extinction 
corrected absolute $Ks$ band magnitudes for both systems, as well as the various measurements that 
were used in their derivation. We previously determined the
brightness of \sco\ in $Ks$ with dedicated observations and it is also detected 
with 2MASS. The quoted error in the magnitude for \sco\ does not reflect the precision of a 
single observation but rather the variability of the source when comparing data from multiple 
epochs. The distance and extinction to \sco\ are fairly well known. \citet{brad99}
determined a distance of 2.8$\pm$0.3 kpc for \sco\ from radio trigonometric parallax.  
\citet{gallagher95} claim an upper limit of $E(B-V)=0.12$ from searches for an 
X-ray scattering halo, while \citet{willis80} favor $E(B-V)=0.35$ from the observed strength of the 
2200\AA\ band. 
In any case, due to the reduced effect of optical extinction in the IR ($A_K=0.114 A_V$; \citealt{card89}),
the extinction at $K$ is at most 0.1 mag which does not have a large impact on the 
derived absolute magnitude.
Estimates for the extinction and distance to \xsix\ are much more uncertain. The extinction to \xsix\ is 
very high, estimated to lie between $A_V=30-50$, based on observed X-ray $N_H$ column densities. 
A distance estimate of $\sim 15$ kpc is given by \citet{christ97}. The final extinction corrected absolute
magnitude range of X1 includes that of \sco\ and thus indicates that the observed $K$ band brightness 
is reasonable for that of a counterpart to \xsix. 

\subsection{\x1702}

For \x1702, one IR source (X) is visible at the edge or partially inside the 0.6\arcsec\
\chandra\ error circle. The source appears somewhat elongated and it is not clear 
whether it would resolve into multiple faint sources at higher resolution.  
An inspection of the individual IR frames is inconclusive in regards to the 
morphology of the source.  
However, {\it Daophot} has no difficulty detecting and fitting its PSF.
Compared to the proposed counterpart for \xsix, the counterpart 
candidate for \x1702\ is quite bright at $Ks= 16.5 \pm 0.07$. A comparison to 
\sco\ is not warranted in this case, since \x1702\ is classified as an Atoll source 
while \sco\ is a member of the Z sources. Apart from distinct differences in their 
X-ray properties, Z and Atoll sources might also indicate particular binary system 
parameters. Atoll sources generally have shorter orbital periods (3--5 hours) than Z sources 
and may differ in the type and/or evolutionary state of the mass donor \citep{band99}. 
Unfortunately, the orbital 
period of \x1702\ is still unknown, so it is difficult to identify an appropriate 
system for comparison. We have chosen a somewhat typical Atoll source for which we 
previously obtained IR photometry, the burster 
X1735$-44$. X1735$-44$ has an orbital period of 4.65~h, an estimated extinction of $A_V=0.7$
and distance of 7-9~kpc \citep{august98}. Table~\ref{t-res} shows that the extinction 
corrected absolute magnitude of \x1702\ is consistent with that of X1735$-$44 given the 
large uncertainties in the input parameters. 

\acknowledgments

The research described in this paper was carried out in part at the
Jet Propulsion Laboratory, California Institute of Technology, and
was sponsored by NASA \chandra\ grants GO2-3044X and GO3-4036X. This
publication makes use of data products from the 2 Micron All Sky
Survey, which is a joint project of the University of Massachusetts
and the Infrared Processing and Analysis Center/California Institute
of Technology, funded by the National Aeronautics and Space
Administration and the National Science Foundation. It also utilized
NASA's Astrophysics Data System Abstract Service and the SIMBAD
database operated by CDS, Strasbourg, France.

\input{tab1.tex}
\input{tab2.tex}
\input{tab3.tex}

\end{document}

%% file: tab1.tex
\begin{deluxetable}{ccccc}
\tablecaption{Log of IR Observations 
\label{t-log} }
\tablecolumns{5}
\tablewidth{0pt}
\tablehead{
\colhead{Date} &
\colhead{Telescope + Instr.} &
\colhead{Filter} &
\colhead{\# Coadds $\times$ Exptime (s)} &
\colhead{\# of Exp.}
}
\startdata
              &                       & {\bf X1624$-$490} &       &    \\
1997 April 28 & CTIO 1.5m + CIRIM     & $J$  & 4 $\times$ 25 & 27 \\
1997 April 26--28 & CTIO 1.5m + CIRIM & $Ks$ & 4 $\times$ 25 & 151 \\
2003 June 19      & ESO NTT + SOFI    & $Ks$ & 10 $\times$ 5 & 17 \\
                  &                   &      &             &     \\
                  &                   & {\bf X1702$-$429} &        &   \\
1995 June 03      & CTIO 1.5m + CIRIM & $Ks$ & 4 $\times$ 15 & 9 \\
2004 July 24      & CTIO 4m + ISPI    & $Ks$ & 3 $\times$ 20 & 9 \\
\enddata
\end{deluxetable}

%% file: tab2.tex
\begin{deluxetable}{ccccccc}
\tablecaption{IR Photometry \label{t-phot}}
\tablewidth{0pt}
\tablehead{
\colhead{Source} &
\colhead{$Ks$} &
\colhead{2MASS Designation}
}
\startdata
\xsix\ &  & \\
A      & 10.623 $\pm$ 0.025 & 16280335-4911535 \\
B      & 12.572 $\pm$ 0.025 & 16280420-4911552\\
C      & 14.81 $\pm$0.07& \\
X1     & 18.31 $\pm$0.1 &  \\
X2     & 18.63 $\pm$0.2 &  \\
\x1702\ & & \\
A      & 12.081 $\pm$0.026 & 17061561-4302042\\
B      & 16.17 $\pm$0.07  & \\
C      & 12.181 $\pm$ 0.026 & 17061714-4301492\\
X      & 16.47 $\pm$0.07  & \\
\enddata
\end{deluxetable}

%% file: tab3.tex
\begin{deluxetable}{cclcccc}
\tabletypesize{\small}
\tablecaption{Results \label{t-res}}
\tablewidth{0pt}
\tablehead{
\colhead{Source} &
\colhead{\chandra\ Coordinates (J2000)} &
\colhead{$m_{Ks}$} &
\colhead{$A_V$} &
\colhead{Distance (kpc)} &
\colhead{$M_{Ks0}$} &
\colhead{Ref.}
}
\startdata
\xsix\  & 16:28:02.825 $-$49:11:54.61 & 18.31 $\pm$0.1  &30--50   &~15        & $-0.99$ to $-3.27$& 1 \\
Sco~X$-$1& \nodata                    & 11.2$\pm$0.3    &0.3--0.9 &2.8$\pm$0.3& $-1.14$       & 2,3,4\\
\x1702\ & 17:06:15.314 $-$43:02:08.69 & 16.47 $\pm$0.07 & ~5     & ~7        & 1.67        & 5 \\
X1735$-$44& \nodata                   & 16.77 $\pm$0.2  & 0.7    & 7$-$9     & 2.46--1.92 & 6,1 \\ 
\enddata
\tablerefs{(1)\citet{christ97} (2)\citet{gallagher95} (3)\citet{willis80} (4)\citet{brad99} (5)\citet{ooster91} (6)\citet{august98} }
\end{deluxetable}